\documentclass[prc,twocolumn,preprintnumbers,superscriptaddress,showpacs,nofootinbib]{revtex4}
\usepackage{bm}
\usepackage{amsmath}
\usepackage{amssymb}
\usepackage[dvips]{graphicx}
\usepackage{times} 
\usepackage{color}

\newcommand{\Psfig}[2]{\includegraphics[width=#1]{#2}}
\newcommand{\rhoB}{\rho_{\scriptscriptstyle B}}

\usepackage{ulem} 

\renewcommand{\em}{\it}

\newcommand{\Slash}[1]{\ooalign{\hfil/\hfil\crcr$#1$}}

\begin{document}
\title{Possibility of $s$-wave pion condensates in neutron stars revisited}
\author{A. Ohnishi}
\author{D. Jido}
\affiliation{
Yukawa Institute for Theoretical Physics, Kyoto University,
Kyoto 606-8502, Japan}
\author{T. Sekihara}
\affiliation{
Department of Physics, Graduate School of Science, Kyoto University,
Kyoto, 606-8502, Japan}
\author{K. Tsubakihara}
\affiliation{
Department of Physics, Faculty of Science,
Hokkaido University, Sapporo 060-0810, Japan}
\date{\today}

\begin{abstract}
We examine possibilities of pion condensation with zero momentum
($s$-wave condensation) in neutron stars
by using the pion-nucleus optical potential $U$
and the relativistic mean field (RMF) models.
We use low-density phenomenological optical potentials
parameterized to fit deeply bound pionic atoms
or pion-nucleus elastic scatterings.
Proton fraction ($Y_p$) and electron chemical potential ($\mu_e$)
in neutron star matter are evaluated in RMF models.
We find that the $s$-wave pion condensation hardly takes place 
in neutron stars and especially has no chance 
if hyperons appear in neutron star matter
and/or $b_1$ parameter in $U$ has density dependence.
\end{abstract}

\maketitle


Pion condensation in neutron stars has a long history of study
from the first suggestion in early 1970s by Migdal~\cite{Migdal:1971cu}
and Sawyer~\cite{Sawyer:1972cq}.
The pion-nucleon interaction is attractive in $p$-wave,
then the main interest along this line was to explore possible appearance 
of pionic excitations with zero energy and {\it finite} momentum,
{\em i.e.} $p$-wave pion condensation~\cite{PionCondensation}
in nuclear matter.
Possibilities of $p$-wave pion condensation in finite nuclei had been
investigated extensively in 1970's and 1980's. Those possibilities were
denied by the non-observation of anomalous angular momentum distribution
in the inelastic excitation of the pionic quantum numbers~\cite{Toki:1979te}.
Possibilities of $p$-wave pion condensation at high densities were also
considered to be improbable based on the universal repulsion assumption,
$g'_{N\Delta} \sim g'_{NN} \sim 0.6$ - $0.8$~\cite{p-con}.
In 1990's, new experiments on the Gamow-Teller giant resonances 
were performed,
and the sum rule value including the $2p-2h$ states was found
to be around 90 \%~\cite{Wakasa:1997zz}, suggesting that the transition
to the $\Delta$ region is weak and $g'_{N\Delta}\sim 0.2$ would be smaller
than $g'_{NN}$~\cite{Suzuki:1998ey}.
In addition, microscopic variational calculation~\cite{APR98} suggests
$\pi^0$ condensation in symmetric nuclear matter at high densities
($\rhoB > 0.2~\mathrm{fm}^{-3}$) generated from the $\Delta$ mediated
three nucleon force.
Thus at present, we cannot completely deny the possibility of pion
condensation in dense matter, and it is necessary to examine
all the ingredients of $\pi N$ and $\pi$-nucleus interactions
with updated experimental and theoretical knowledge.

The study of the in-medium pion properties,
especially the $s$-wave pion-nucleus interaction,
has been recently developed in experiment.
Precise observations of deeply bound atomic states of $\pi^{-}$
in Pb and Sn isotopes~\cite{Gilg:1999qa,Itahashi:1999qb,Suzuki:2002ae,Kienle,PionicAtom} 
and low energy pion-nucleus elastic scattering~\cite{PionScattering}
provide us with detailed information of 
density dependent optical potentials at low densities. 
Theoretical calculations of in-medium pion self-energy have also experienced 
much progress 
based on chiral dynamics~\cite{Thorsson:1995rj,Meissner:2001gz,Kolomeitsev:2002gc,Doring:2007qi}. 

Together with these developments,
it may be interesting to revisit pion condensation at high densities, such as 
in neutron stars.
Motivated by the recent progress in the $s$-wave pion-nucleus interaction,
we concentrate on the study of pion condensation with zero momentum
($s$-wave condensation), for simplicity.  
Even such a limited investigation would make progress of our 
understanding of dense matter. 
The $s$-wave pion condensation takes place in neutron stars
with nuclear matter instability 
where the transition $n \to p\pi^-$ becomes energetically
possible~\cite{Glendenning}. This happens with 
$\mu_n - \mu_p \ge E_{\pi}$, where $\mu_n$ and $\mu_p$ are
the neutron and proton chemical potentials, respectively, and 
$E_{\pi}$ is the $\pi^{-}$ energy at rest. Due to 
$\beta$ equilibrium under the charge neutral and neutrino-less 
conditions, $\mu_n - \mu_p$ 
can be written as the electron chemical 
potential $\mu_e=\mu_n - \mu_p$. 
Since $nn$ interaction is more repulsive than $pn$,
$\mu_n$ is pushed up then $\mu_e$ increases compared to Fermi gas value
in neutron rich matter.
For example, relativistic mean field (RMF) models
suggest~\cite{Glendenning,IOTSY} that $\mu_e$ largely exceeds
the in-vacuum $\pi^{-}$ mass at 
nuclear density
$\rhoB=(1\sim 5) \rho_{0}$ in neutron stars,
where $\rho_{0}$ is the saturation density.

In this paper, 
we examine whether the condition for the $s$-wave $\pi^{-}$ condensation,
$E_{\pi} = \mu_{e}$,
is satisfied in neutron stars for $E_{\pi}$ and $\mu_{e}$
obtained in our present knowledge of 
low density pion optical potentials and equation of state (EOS) of neutron star matter.
For $E_{\pi}$, we use various pion optical potentials fitted so as to 
reproduce the pionic atom~\cite{Kienle,Tauscher:1971,B2,SM1,E1,Nieves:1993ev}
and $\pi$-nucleus elastic scattering data~\cite{PionScattering}.
{Although the fitted optical potentials  in normal nuclei 
may not be extrapolated to higher density and/or highly asymmetric 
nuclear matter,
it is interesting to examine the present status and to think about next steps.}
For $\mu_{e}$, we adopt the results calculated in RMF
with several parameter sets~\cite{TM1,NL1,NL3,TO_2007},
which explain the bulk properties
of nuclei such as the binding energy and the charge radius in a wide mass range.
We also use proton fractions ($Y_p$) evaluated with RMF
for calculating pion optical potentials\footnote{A preliminary study has been done along the same line
for limited combinations of pion potentials
and an RMF parameter set~\cite{IOTSY}.}.
There are several theoretical works on the $s$-wave pion condensation 
at higher densities~\cite{higher},
while the connection to low density phenomena observed in experiments
is not clear yet.
Since we concentrate on the possibility of the $s$-wave $\pi^{-}$ condensate,
we do not consider the double pole condition for $\pi^{-} \pi^{+}_{s}$ pair 
creation, which takes place with finite momentum.

The $\pi^{-}$ energy in uniform matter may be evaluated by
$E_\pi=\sqrt{m_\pi^2+2m_\pi{U}}$ ($\mathbf{p}=0$) with
the real part of the {\it energy-independent} potential $U$
{based on Ericson-Ericson parameterization~\cite{Ericson}:} 
\begin{equation}
U=-\frac{2\pi}{m_\pi}\left[
	\left(1+\epsilon \right)\left(b_0\rhoB
	+b_1\delta\rho\right)
	+\left(1+\frac{\epsilon}{2}\right)B_0^{\rm Re}\,\rho^{(2)} 
	\right] \nonumber
\end{equation}
with $\epsilon = m_{\pi}/M_{N}$, $\rhoB=\rho_n+\rho_p$, 
$\delta\rho=\rho_n-\rho_p=\rhoB(1-2Y_p)$ and 
a squared density $\rho^{(2)}$ defined below.
This potential is related to the pion self-energy
via $\Sigma_\pi=2m_\pi(U+iW)$ with an imaginary potential $W$.
The $s$-wave $\pi{N}$ potential parameters  ($b_0, b_1, B_0^{\rm Re}$) 
was determined with special care
from precise measurements of pionic atom 
and pion-nucleus scattering data.

In Table~\ref{Table:PiPot}, we summarize the parameter sets
adopted here. The upper 6 sets were 
determined from the pionic atom data
and the lower two from pion-nucleus scattering.
For the former parameter sets except NOG, 
$b_{0}=\tilde b_{0}$ and $\rho^{(2)} = \rhoB^{2}$
are used,
whereas, for the latter, double scattering modifications were explicitly 
included by 
$b_{0} = \tilde b_{0} - 3(\tilde b_{0}^{2}+2 \tilde b_{1}^{2}) k_{F}/(2\pi) $
with $k_{F} = (3 \pi ^{2} \rhoB /2)^{1/3}$, 
and $\rho^{(2)} = \rhoB^{2} - \delta\rho^{2}$ is used.
For NOG, $b_{0}=\tilde b_{0} + \delta b_{0}
- 3(1+\epsilon)(\tilde b_{0}^{2}+2 \tilde b_{1}^{2}) k_{F}/(2\pi) $ with $\delta b_{0}=-0.0053 m_{\pi}^{-1}$.
In the parameter sets of KY and F-W, $b_{1}$ 
is assumed to have density dependence through 
$b_1 = \tilde b_1/(1-\alpha\rhoB/\rho_0)$~\cite{MedFpi,Jido:2000bw,Kolomeitsev:2002gc} with finite $\alpha$ and $\rho_0=0.17~\mathrm{fm}^{-3}$.
This is a consequence of the pion wave function renormalization
associated with energy dependence of the optical potential~\cite{Jido:2000bw,Kolomeitsev:2002gc}
and the renormalization was performed at $E_{\pi}=m_{\pi}$.
For more realistic calculations in dense nuclear matter, the wave function 
renormalization should be done at the in-medium pion mass
and other parameters should be also renormalized.

\begin{table}
\caption{Pion potential parameters. The upper four sets by 
the pionic atom are taken from Ref.~\cite{Itahashi:1999qb},
in which $b_{0}$ and $B_{0}^{\rm Im}$ were readjusted to reproduce 
the recent data of 
the deeply bound $\pi^{-}$ states in Pb with fixing the other parameters
as the original values given in Ref.~\cite{Tauscher:1971} for T,
\cite{B2} for BFG, \cite{SM1} for SM and \cite{E1} for ET.}
\label{Table:PiPot}
\begin{tabular}{ll|ccccc}
\hline\hline
&parameter&$\tilde b_0$          &$\tilde b_1$ &$B_0^{\rm Re}$ & $\alpha$ &\\
system& set&$(m_\pi^{-1})$ &$(m_\pi^{-1})$ &$(m_\pi^{-4})$   & & \\
\hline
&T 
                       & $-0.034 \phantom{0}$  & $-0.078 \phantom{0}$   
                       & $\phantom{-} 0 \phantom{.000}$    & $0$      \\
pionic &BFG 
& $-0.025 \phantom{0}$  & $-0.085 \phantom{0}$   
                       & $-0.021$  & $0$       \\
atom &SM 
                       & $-0.027 \phantom{0}$  & $-0.12 \phantom{00}$    
                       & $\phantom{-} 0 \phantom{.000}$ & $0$ \\
&ET 
    & $-0.020 \phantom{0}$  & $-0.0873$  & $-0.049$  & $0$       \\
&NOG \protect{\cite{Nieves:1993ev}}&  $-0.013 \phantom{0}$&$-0.105\phantom{0}$  &$\phantom{-} 0 \phantom{.000}$ & $0$\\
&KY\protect{\cite{Kienle}}
                       & $-0.0233$ &$-0.1473$   & $-0.019$  & $0.367$ \\
\hline
pion-nucleus &F-C\protect{\cite{PionScattering}}
                       & $-0.009 \phantom{0}$  & $-0.114 \phantom{0}$  & $-0.040$   & $0$ \\
scattering &F-W\protect{\cite{PionScattering}}
                       & $-0.009 \phantom{0}$  & $-0.081 \phantom{0}$  & $-0.040$   & $0.391$ & \\
\hline
\hline
\end{tabular}
\end{table}

%
{The phenomenological potentials are determined  with a fixed pion energy.}
The energy dependence of the optical potential can be estimated 
by theoretical calculations.
The $s$-wave in-medium pion self-energy $\Sigma_{\pi}(E_{\pi})$, 
equivalent to the optical 
potential, was derived
based on the chiral perturbation theory in Ref.~\cite{Thorsson:1995rj}
within a linear density approximation.
The in-medium pion mass is obtained by solving $m_{\pi}^{\ast 2}-m_{\pi}^{2}
- \Sigma(m_{\pi}^{\ast}) = 0$, which automatically takes account of the 
wave function renormalization. Here we use the following self-energy, 
abbreviated as MOW~\cite{Meissner:2001gz},
\begin{equation*} 
  \Sigma (m_{\pi}^{\ast})  =  c_{1} \frac{4 \rhoB}{f^{2}} m_{\pi}^{2} - 
  \frac{2 \rhoB}{f^{2}} m_{\pi}^{\ast 2} 
  \left(c_{2} + c_{3} - \frac{g_{A}^{2}}{8m_{N}} \right)
  + \frac{m_{\pi}^{\ast} \delta \rho}{2f^{2}}
  \ ,
\end{equation*}
with $c_{1}=-0.81$, $c_{2}=3.20$, $c_{3}=-4.66$ in units of GeV$^{-1}$ 
and $f=88$ MeV.
We also consider the $\pi^{-}$ 
self-energy calculated by an in-medium chiral perturbation
theory in ${\cal O}(p^{5})$ discussed in Ref.~\cite{Kolomeitsev:2002gc} (KKW),
which reproduces well the energies and widths of deeply bound $\pi^{-}$ atomic 
states in Pb.

\begin{figure}
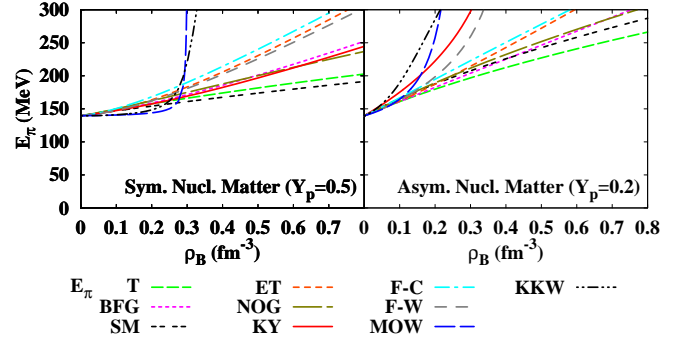

\centerline{
\Psfig{9cm}{Epi.eps}
}
\caption{(Color online)
$E_\pi$ in symmetric ($Y_p=0.5$, upper)
and asymmetric ($Y_p=0.2$, lower) nuclear matter.
}
\label{Fig:Epi}
\end{figure}

\begin{figure}
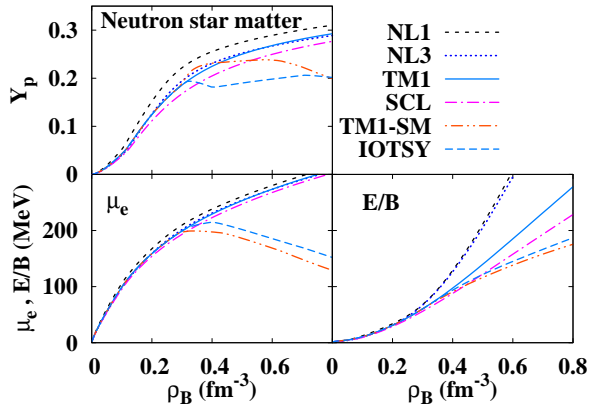

\centerline{
\Psfig{8.0cm}{RMF.eps}
}
\caption{(Color online)
RMF results of 
proton fraction ($Y_p=\rho_p/\rhoB$, upper panel), 
electron chemical potential ($\mu_e$, lower left panel)
and energy per baryon ($E/B$, lower right panel).
}
\label{Fig:RMF}
\end{figure}

Let us see the model dependence of the pion energy $E_\pi$.
In Fig.~\ref{Fig:Epi}, 
we show $E_\pi$ in the cases of symmetric ($Y_p=0.5$)
and asymmetric ($Y_p=0.2$) nuclear matter.
The proton fraction $Y_p =0.2$ is a typical value
in neutron star matter obtained in RMF, as shown later in Fig.~\ref{Fig:RMF}.
The negative signs of the coefficients ($b_0, b_1, B_0^{\rm Re}$)
imply that $\pi^-$ feels repulsive potential in nuclear matter.
The phenomenological pion potentials in symmetric nuclear matter
agree well with each other at low densities below $\rho_0$,
whereas, in asymmetric nuclear matter with $Y_p=0.2$,
we have $50$-$100$ MeV ambiguities at $\rhoB \sim \rho_0$.
In order to 
fix the large ambiguity of the potentials in asymmetric nuclear matter, 
it is very interesting to obtain pionic atom and scattering data
in neutron rich nuclei~\cite{RIKEN}.

\begin{table*}[t]
\caption{RMF parameters. In SCL, 
$g_3$ and $g_4$ are from the expansion of $f_\mathrm{SCL}$.}
\label{Table:RMF}
\begin{tabular}{l|rrrrrrrrr}
\hline\hline
&$g_{\sigma{N}}$
&$g_{\omega{N}}$
&$g_{\rho{N}}$
&$g_3$(MeV)
&$g_4$
&$c_\omega$
&$m_\sigma$(MeV)
&$m_\omega$(MeV)
&$m_\rho$(MeV)
\\
\hline
NL1\protect{\cite{NL1}}
&10.138&13.285&4.976&2401.9&-36.265&0&492.25&795.359&763 \\
NL3\protect{\cite{NL3}}
&10.217&12.868&4.474&2058.35&-28.885&0&508.194&782.501&763\\
TM1\protect{\cite{TM1}}
&10.0289&12.6139&4.6322&1426.466&0.6183&71.3075&511.198&783&770 \\
SCL\protect{\cite{TO_2007}}
&10.08&13.02&4.40&1255.88&13.504&200&502.63&783&770\\
\hline\hline
\end{tabular}
\end{table*}

RMF models have been developed to describe bulk properties of nuclei
and nuclear matter with the mean field via the meson fields.
We here adopt the RMF models,
NL1~\cite{NL1}, NL3~\cite{NL3}, TM1~\cite{TM1}, SCL~\cite{TO_2007},
having the Lagrangian in the following form,
\begin{align}
{\cal L}=&{\cal L}_\mathrm{free}
+\bar{\psi}\left[
	g_\sigma\sigma-g_\omega\Slash{\omega}-g_\rho\tau_z\Slash{\rho}
	\right]\psi
+\frac{c_\omega}{4}\omega^4-V_\sigma(\sigma)
\nonumber
,\\
V_\sigma=&
\begin{cases}
\frac13 g_3\sigma^3+\frac14 g_4\sigma^4 & (\mbox{NL1, NL3, TM1}) \\
- a_\sigma f_\mathrm{SCL}(\sigma/f_\pi) & (\mbox{SCL})
\end{cases}
\ , \nonumber
\end{align}
where
$\psi$, $\sigma$, $\omega$, $\rho$ represent
nucleon and $\sigma$-, $\omega$- and $\rho$-meson fields, respectively,
and $f_\mathrm{SCL}(x)=\log(1-x)+x+x^2/2$.
The model parameters are summarized in Table~\ref{Table:RMF}.
These RMF models describe the binding energies of
heavy semi-double magic nuclei well,
and are expected to give reasonable EOS of nuclear matter.
We have solved the $\beta$ equilibrium condition in cold neutron star matter,
\begin{align}
\mu_e &= \mu_n - \mu_p\ ,\quad
\rho_e = \rho_p\ ,
 \label{Eq:mu_e}
\\
\mu_{n,p} &=
 \sqrt{{M_N^*}^2+k_F^2}+g_\omega\omega \mp g_\rho \rho 
\label{Eq:mu_N}\ ,
\end{align}
where $M_N^*=M_N-g_\sigma\sigma$ represents the effective mass of nucleon.
As shown in Fig.~\ref{Fig:RMF},
calculated values of $E/B$, $Y_p$ and $\mu_e$ in neutron star matter 
are consistent 
at low densities ($\rhoB<\rho_0$),
since meson-baryon coupling constants are
well determined
by the binding energies of heavy-nuclei.
Significant differences are found in $E/B$ at higher densities,
where the mesons have large expectation values
and the self-interaction terms ($\sigma^3, \sigma^4, \omega^4$)
contribute to $E/B$ considerably.
While we have small differences in $Y_p$ and $\mu_e$,
the model dependence is smaller compared with those 
in $E/B$ and $E_\pi$.
As we can see from Eq.(2), $\mu_n-\mu_p$ is modified
from the Fermi gas value with $M^*$
by the $\rho$ meson,
whose coupling with nucleons is well constrained by nuclear binding energies,
and higher order terms of the $\rho$ meson are not included in the RMF models
under consideration.
As a result, model dependence of the isospin dependent potential,
$g_\rho\rho$, is only around 10 MeV at $\rhoB=0.8~\mathrm{fm}^{-3}$.
In Fig.~\ref{Fig:RMF}
we also show the results of some RMF models
including hyperons (TM1-SM~\cite{Schaffner} and IOTSY~\cite{IOTSY}).
With hyperons, the proton fraction and electron chemical potential
significantly decrease.

\begin{figure}
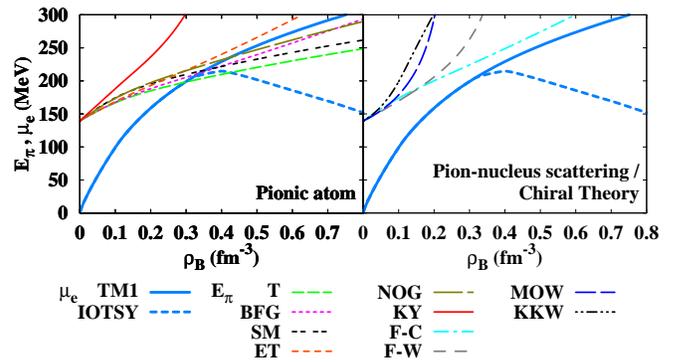

\centerline{
\Psfig{9cm}{Comp.eps}
}
\caption{(Color online)
Electron chemical potentials in RMF models
and pion energy in neutron stars.
}
\label{Fig:Comp}
\end{figure}

Now let us compare the electron chemical potential $\mu_{e}$ and the 
pion energy $E_{\pi}$ as functions of $\rhoB$ (Fig.~\ref{Fig:Comp}). 
The RMF results of TM1 are adopted for the proton fraction ($Y_p$)
to evaluate $E_\pi$.
Results with hyperons denoted by IOTSY are also shown.
The left panel of Fig.~\ref{Fig:Comp} shows 
the comparison of $\mu_e$ and $E_\pi$ obtained from the pionic atom data, 
namely, Tauscher (T),
Batty-Friedman-Gal (BFG),
Seki-Masutani (SM),
Ericson-Tauscher (ET) and
Kienle-Yamazaki (KY).
In these potentials, density dependence of the 
potential parameters is not taken into account except for KY.
The right panel of Fig.~\ref{Fig:Comp} compares $\mu_{e}$
with $E_\pi$ 
obtained from the pion-nucleus scattering data
and the theoretical calculations.

We find that $E_\pi$ obtained from the potentials 
with density-independent $b_{1}$  are very close
to  $\mu_e$ at high densities $\rhoB > 0.3~\mathrm{fm}^{-3}$,
and in further dense nuclear matter, $\mu_e$ exceeds $E_\pi$
obtained with some of the parameter sets.
We could have possibility for the $s$-wave pion condensation to take
place in dense neutron star matter.
However, since inclusion of hyperons makes $\mu_{e}$ suppressed, 
$E_{\pi}$ is found to be larger than $\mu_{e}$ (IOTSY) in
most cases.
Thus the $s$-wave pion condensation would not take place, if hyperons could
participate in neutron star matter.

We note that even more attractive hyperon potentials make $\mu_e$ smaller
(TM1-SM).
Also in non-relativistic variational treatments~\cite{FP81,APR98}, 
symmetry energy and $\mu_e$ are generally smaller at $\rhoB > \rho_0$
than in relativistic models.
In the case of the density-dependent $b_{1}$, which is a consequence of 
the renormalization of the pion wave function
and is required to explain the pionic
atom data of Sn isotopes~\cite{Kienle},
the pion self-energies are more repulsive. 
Nevertheless, it is important to note that, as already mentioned,
the renormalization of the wave function has been done only for the $b_{1}$
parameter in linear $\delta \rho$.
Thus, for more quantitative discussion,
it is necessary to improve the phenomenological and theoretical
pion optical potentials in a consistent way,
for instance as done in Ref.~\cite{Doring:2007qi}.
It is also desired to include the effects of short-range and tensor
correlations on $\mu_e$ under $\beta$-equilibrium~\cite{APR98}
in relativistic frameworks~\cite{RBHF}.

In summary, 
we have discussed the in-medium pion energy in the context of
possibility for the $s$-wave pion condensation to take place in neutron stars.
We have compared the in-medium 
pion energies determined from pionic atom or pion-nucleus
scattering data
with the electron chemical potential evaluated in relativistic mean field (RMF)
models, using the RMF result of the proton fraction.
With our present limited knowledge of the in-medium pion properties
obtained in experiments, we could conclude that the $s$-wave
pion condensation would not take place in neutron stars with hyperons.
It is certainly necessary to investigate in-medium pion self-energy 
theoretically in more elaborated prescription to go beyond nuclear
density. Especially energy dependence of the pion self-energy 
should be treated in more proper ways for higher densities. 
At the same time, experimental observations of pionic atoms
and scattering in neutron rich nuclei are essential  
to fix ambiguities in the pion optical potentials in asymmetric nuclear matter. 
Precise knowledge of pion self-energies at high density is also important to
study finite temperature process 
such as black hole formations,
where the temperature can be as large as $T=70~\mathrm{MeV}$~\cite{BH}.
At such high temperatures, pion contribution could be significant depending on the 
in-medium pion mass. 
We acknowledge S.~Hirenzaki, T.~Kunihiro, K.~Sumiyoshi
and K.~Yazaki for useful discussions.
This work was supported in part by
the Grant-in-Aid for Scientific Research
from MEXT and JSPS
under the grant numbers,
   17070002,	
   19540252,		
   20028004    	
   and 20540273, 
and the Yukawa International Program for Quark-hadron Sciences (YIPQS).
KT also thanks JSPS for the fellowship (20$\cdot$4326).




\begin{thebibliography}{99}
\bibitem{Migdal:1971cu}
  A.~B.~Migdal,
  Zh.\ Eksp.\ Teor.\ Fiz.\  {\bf 61} (1971) 2209
  [Sov. Phys. JETP {\bf 36}, 1052 (1973)].

\bibitem{Sawyer:1972cq}
  R.~F.~Sawyer,
  Phys.\ Rev.\ Lett.\  {\bf 29}, 382 (1972).

\bibitem{PionCondensation}
For example,
A. B. Migdal,
Rev. Mod. Phys. {\bf 50}, 107 (1978);
%
S. O. Baeckman and W. Weise,
Phys. Lett. {\bf B55}, 1 (1975);
%
G. Baym and E. Flowers,
Nucl. Phys. {\bf A222}, 29 (1974).
%
  
\bibitem{Toki:1979te}
  H.~Toki and W.~Weise,
  Phys.\ Rev.\ Lett.\  {\bf 42}, 1034 (1979). 

\bibitem{p-con}
  W.~H.~Dickhoff, A.~Faessler, J.~Meyer-Ter-Vehn and H.~Muther,
  Phys.\ Rev.\  C {\bf 23}, 1154 (1981);
  W.~H.~Dickhoff, A.~Faessler, H.~Muther and J.~Meyer-Ter-Vehn,
  Nucl.\ Phys.\  A {\bf 368}, 445 (1981);
  E.~Oset and A.~Palanques-Mestre,
  Phys.\ Rev.\  C {\bf 30}, 366 (1984);
  T.~Kunihiro, T.~Takatsuka, R.~Tamagaki and T.~Tatsumi,
  Prog.\ Theor.\ Phys.\ Suppl.\  {\bf 112}, 123 (1993).


\bibitem{Wakasa:1997zz}
 T.~Wakasa {\it et al.},
 Phys.\ Rev.\  C {\bf 55}, 2909 (1997).
 
 
\bibitem{Suzuki:1998ey}
 T.~Suzuki and H.~Sakai,
 Phys.\ Lett.\  B {\bf 455}, 25 (1999).

\bibitem{APR98}
 A.~Akmal, V.~R.~Pandharipande and D.~G.~Ravenhall,
 Phys.\ Rev.\  C {\bf 58}, 1804 (1998).


\bibitem{Gilg:1999qa}
H. Gilg {\it et al.},
Phys. Rev. {\bf C62}, 025201 (2000);

\bibitem{Itahashi:1999qb}
  K.~Itahashi {\it et al.},
  Phys.\ Rev.\  C {\bf 62}, 025202 (2000).

\bibitem{Suzuki:2002ae}
K. Suzuki {\it et al.},
Phys. Rev. Lett. {\bf 92}, 072302 (2004).

\bibitem{Kienle}
P. Kienle and T. Yamazaki,
Prog. Part. Nucl. Phys. {\bf 52}, 85 (2004).

\bibitem{PionicAtom}
See also, S. Hirenzaki, H. Toki and T. Yamazaki,
Phys. Rev. {\bf C44}, 2472 (1991).

\bibitem{PionScattering}
  E.~Friedman {\it et al.}
  Phys.\ Rev.\ Lett.\ {\bf 93} 122302 (2004);
E. Friedman {\it et al.},
Phys. Rev. {\bf C72}, 034609 (2005).


\bibitem{Thorsson:1995rj}
V. Thorsson and A. Wirzba,
Nucl. Phys. {\bf A589}, 633 (1995).

\bibitem{Meissner:2001gz}
  U.~G.~Meissner, J.~A.~Oller and A.~Wirzba,
  Ann.\ Phys.\ (N.Y.)  {\bf 297}, 27 (2002).
  
  \bibitem{Kolomeitsev:2002gc}
E.~E.~Kolomeitsev, N.~Kaiser and W.~Weise,
Phys.\ Rev.\ Lett.\ {\bf 90}, 092501 (2003).

\bibitem{Doring:2007qi}
M. Doring and E. Oset,
Phys. Rev. {\bf C77}, 024602 (2008).


\bibitem{Glendenning}
  N. K. Glendenning,
  "Compact Stars: Nuclear Physics, Particle Physics and General Relativity"
  (Springer-Verlag, Berlin, 2000),
  and references therein.
%
\bibitem{IOTSY}
  C.~Ishizuka, A.~Ohnishi, K.~Tsubakihara, K.~Sumiyoshi and S.~Yamada,
  J.\ Phys.\ G {\bf 35}, 085201 (2008).
 

\bibitem{Tauscher:1971}
  L.~Tauscher, 
  in Proceedings of the International Seminar on $\pi$-Meson Nucleus 
  Interaction Strasbourg 1971, CNRS-Strasbourg (unpublished), p.~45.

\bibitem{B2}
  C.~J.~Batty, E.~Friedman and A.~Gal, Nucl. Phys. A {\bf 402}, 411 (1983). 

\bibitem{SM1}
  R. Seki and K. Masutani, Phys.\ Rev.\  C {\bf 27}, 2799 (1983).

\bibitem{E1}
  T.E.O. Ericson and L. Tauscher, Phys. Lett. B {\bf 112}, 425 (1982).

\bibitem{Nieves:1993ev}
  J.~Nieves, E.~Oset and C.~Garcia-Recio,
  Nucl.\ Phys.\  A {\bf 554}, 509 (1993).

\bibitem{NL1}
  P.~G.~Reinhard, M.~Rufa, J.~Maruhn, W.~Greiner and J.~Friedrich,
  Z.\ Phys.\  A {\bf 323}, 13 (1986);
  S.~J.~Lee {\it et al.},
  Phys.\ Rev.\ Lett.\  {\bf 57}, 2916 (1986).
  
\bibitem{NL3}
  G.~A.~Lalazissis, J.~Konig and P.~Ring,
  Phys.\ Rev.\  C {\bf 55}, 540 (1997).

\bibitem{TM1}
  Y.~Sugahara and H.~Toki,
  Nucl.\ Phys.\  A {\bf 579}, 557 (1994).

\bibitem{TO_2007}
  K.~Tsubakihara and A.~Ohnishi,
  Prog.\ Theor.\ Phys.\  {\bf 117}, 903 (2007).
  
  
\bibitem{higher}
  L.y.~He, M.~Jin and P.f.~Zhuang,
  Phys.\ Rev.\  D {\bf 71}, 116001 (2005);
%
  D.~Ebert and K.~G.~Klimenko,
  Eur.\ Phys.\ J.\  C {\bf 46}, 771 (2006);
%
  H.~Abuki, R.~Anglani, R.~Gatto, M.~Pellicoro and M.~Ruggieri,
  Phys.\ Rev.\  D {\bf 79} (2009) 034032.
%
  D.~T.~Son and M.~A.~Stephanov,
  Phys.\ Atom.\ Nucl.\  {\bf 64}, 834 (2001);
%
  H.~J.~Lee, B.~Y.~Park, D.~P.~Min, M.~Rho and V.~Vento,
  Nucl.\ Phys.\  A {\bf 723}, 427 (2003);
%
  J.~O.~Andersen,
  Phys.\ Rev.\  D {\bf 75}, 065011 (2007).


\bibitem{Ericson}
M.~Ericson and T.E.O.~Ericson, Ann.\ Phys.\ (N.Y.) {\bf 36}, 323 (1966).
 


\bibitem{MedFpi}
W.~Weise,
Acta Phys.\ Polon.\ {\bf B31}, 2715 (2000);
Nucl.\ Phys.\ {\bf A690}, 98c (2001).

\bibitem{Jido:2000bw}
D.~Jido, T.~Hatsuda and T.~Kunihiro,
Phys.\ Rev.\ {\bf D63}, 011901 (2001); 
  Phys.\ Lett.\  B {\bf 670}, 109 (2008).
  
\bibitem{RIKEN}
For example, K.~Itahashi, RIBF proposal, NP0802-RIBF~54, (2008).

\bibitem{Schaffner}
 J. Schaffner and I. Mishustin,
 Phys. Rev C \textbf{53}, 1416 (1996).
%

\bibitem{FP81}
 B.~Friedman and V.~R.~Pandharipande,
 Nucl.\ Phys.\  A {\bf 361}, 502 (1981);
\bibitem{RBHF}
  R.~Brockmann and R.~Machleidt,
  Phys.\ Rev.\  C {\bf 42}, 1965 (1990).
\bibitem{BH}
  K.~Sumiyoshi, S.~Yamada, H.~Suzuki and S.~Chiba,
  Phys.\ Rev.\ Lett.\  {\bf 97}, 091101 (2006);
%
  K.~Sumiyoshi, C.~Ishizuka, A.~Ohnishi, S.~Yamada, H.~Suzuki, (2008),
  Astrophys.\ J.\ Lett.\ {\bf 690}, L43 (2009).
\end{thebibliography}
\end{document}